\PassOptionsToPackage{numbers}{natbib}
\documentclass{article}
\usepackage[preprint]{neurips_2025}
\usepackage[utf8]{inputenc} 
\usepackage[T1]{fontenc}    
\usepackage{hyperref}       
\usepackage{url}            
\usepackage{booktabs}       
\usepackage{amsfonts}       
\usepackage{nicefrac}       
\usepackage{microtype}      
\usepackage{xcolor}         
\usepackage{amsmath}        
\usepackage{array}          
\usepackage{longtable}      
\usepackage{enumitem}       
\usepackage{graphicx}       
\usepackage{floatrow}

\title{Typing Reinvented: Towards Hands-Free Input via sEMG}

\author{
  Kunwoo Lee,
  Dhivya Sreedhar,
  Pushkar Saraf,
  Chaeeun Lee,
  Kateryna Shapovalenko \\
  Carnegie Mellon University, Pittsburgh, PA 15213 \\
  \texttt{\{kunwool, dsreedha, psaraf, chaeeunl\}@andrew.cmu.edu} \\
  \texttt{kshapova@alumni.cmu.edu}
}

\begin{document}

\maketitle

\begin{abstract}
We explore surface electromyography (sEMG) as a non-invasive input modality for mapping muscle activity to keyboard inputs, targeting immersive typing in next-generation human–computer interaction (HCI). This is especially relevant for spatial computing and virtual reality (VR), where traditional keyboards are impractical. Using attention-based architectures, we significantly outperform the existing convolutional baselines, \textbf{reducing online generic CER from 24.98\% → 20.34\% and offline personalized CER from 10.86\% → 10.10\%}, while remaining fully causal. We further incorporate a lightweight decoding pipeline with language-model–based correction, demonstrating the feasibility of accurate, real-time muscle-driven text input for future wearable and spatial interfaces.
\end{abstract}

\section{Introduction}

Since the invention of the typewriter in 1868, the basic design of typing devices (machines with buttons representing characters) has remained unchanged. As keyboards are crucial for translating human intent into computer input, it’s worth reevaluating their relevance, especially as new computing formats, like Apple Vision Pro’s spatial computing, emerge.

In this new era of computing, traditional keyboards may become obsolete, yet the instinct to press keys while typing remains deeply embedded in how we interact with technology. For instance, Vision Pro’s gaze-based typing proves impractical for tasks like document creation. This highlights the need for virtual keyboards that preserve the familiar act of typing without relying on physical hardware or a desk-enabling input in mid-air through a floating interface. Our approach leverages sEMG to map muscle signals to virtual keystrokes, which is anticipated to be suitable in this context.

With this system, we aim to replicate the natural experience of typing by capturing fine-grained muscle activity, preserving users’ cognitive and physical expectations. To validate our approach, we target state-of-the-art performance on the emg2qwerty dataset using modern neural architectures such as Transformers and Conformers. We evaluate model accuracy using character error rate (CER), benchmark against existing baselines, and explore latency and inference speed to assess the system’s viability for real-time mid-air typing.

\section{Related Work and Background}

Early work on EMG-driven text input covered a range of topics and produced encouraging, but ultimately limited results, all constrained by the modest size of their data.  An early brain-computer-interface study combined eye tracking with EMG signals to control a virtual keyboard; it showed that users could reliably select characters with mean accuracies above 90 \%, yet the experiment involved only five participants and roughly 1000 keystrokes in total, forcing the authors to rely on handcrafted thresholds rather than trainable models \citet{ref1}. A differential-EMG investigation of forearm muscles during sequential key presses explored how muscle activation varies with tempo and finger order, reporting clear, class-separable activation patterns and classification accuracies exceeding 85 \%, but the study drew on fewer than 15 subjects and just minutes of data per task, precluding the use of data-hungry deep networks \citet{chong2015differential}. Prototypes such as Air Keyboard and MyoKey tackled mid-air typing directly: Air Keyboard achieved about 7 words per minute with heuristic gesture mapping, while MyoKey reached 9 WPM using a shallow CNN; both systems, however, were trained on no more than a few thousand keystrokes from 10 users, making overfitting a central concern and limiting vocabulary size \citet{gaba2016air}. Across these studies, the common thread is clear: despite promising proof-of-concept results, each investigation relied on ad-hoc datasets—typically under 20 minutes of sEMG per participant, rendering them inadequate for modern high-capacity sequence models such as Transformers or Conformers that thrive on large, diverse corpora.

As an effort to push the domain forward, Meta released two large-scale datasets: \href{https://arxiv.org/abs/2410.20081}{emg2qwerty} and \href{https://github.com/facebookresearch/emg2pose}{emg2pose}.
\begin{itemize}
    \item \textbf{emg2qwerty}: Contains non-invasive EMG signals collected from 108 users over 1335 sessions, over 346 hours, making it the largest public dataset of its kind. \citet{sivakumar2024emg2qwertylargedatasetbaselines}
    \item \textbf{emg2pose}: Focuses on hand-based sEMG for estimating poses. The data set includes 193 users, 370 hours of data, and 29 stages featuring various gestures, offering a scale comparable to vision-based hand pose datasets. It contains 2kHz, 16-channel sEMG recordings paired with high-quality hand pose labels obtained from a 26-camera motion capture system.  \citet{salter2024emg2poselargediversebenchmark}
\end{itemize}

For this project, we focus on the emg2qwerty dataset, with plans to incorporate the emg2pose dataset in the future to generate complementary virtual hand outputs for a more immersive environment. The size and richness of the emg2qwerty dataset have opened up new opportunities for researchers to develop innovative models and tools that advance the field. However, as a relatively new resource, it has yet to be fully explored, and no project have leveraged its full potential beyond baseline. Our work aims to address this gap by exploring advanced architectures, such as Transformers and Conformers, to surpass the current baseline and achieve state-of-the-art (SOTA) performance in sEMG-to-keyboard input mapping.

\section{Dataset}

\begin{figure}
    \centering
    \includegraphics[width=0.8\linewidth]{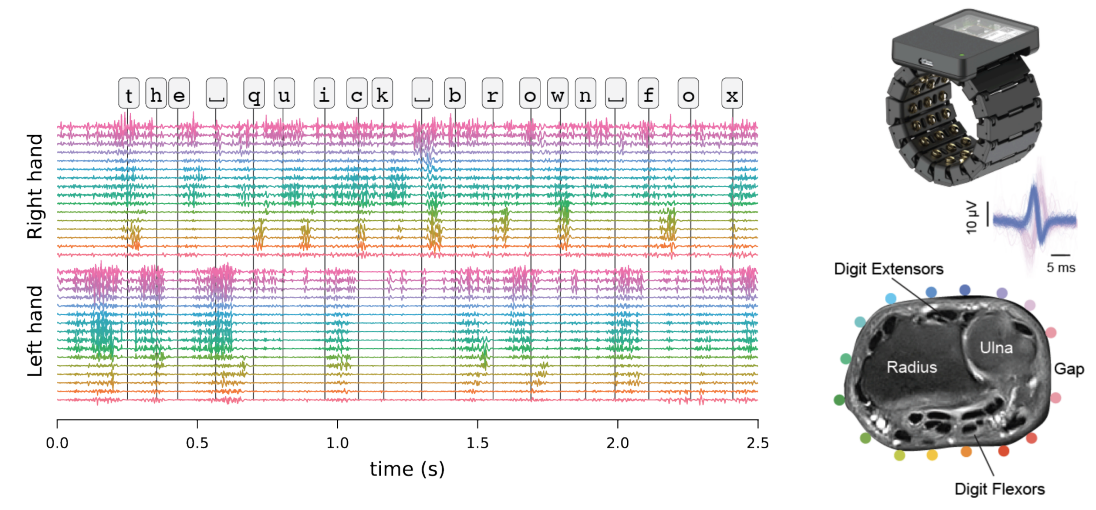}
    \caption{Dataset overview (\citet{emg2qwerty})}
    \label{fig:emg}
\end{figure}

The dataset in \citet{emg2qwerty} is currently the largest dataset for sEMG labeled by keyboard input as in Fig.~\ref{fig:emg}. It comprises 1135 sessions, 108 users, and 346 hours of recordings. From this dataset, \citet{emg2qwerty} demonstrated that with minimal investment, high performance can be achieved by leveraging existing architectures from related domains. Specifically, Time Depth Separable Convolution (TDS), a model popular in Automatic Speech Recognition (ASR) was used and achieved CER below 10\%. This is motivating for us as there may be large room of improvement, potentially paving the way for the next generation of typing experiences.

\begin{figure}[!ht]
    \centering
    \begin{floatrow}
      \ffigbox[\FBwidth]{\caption{Inter-channel correlation}\label{fig:icc}}{%
        \includegraphics[width=0.8\linewidth]{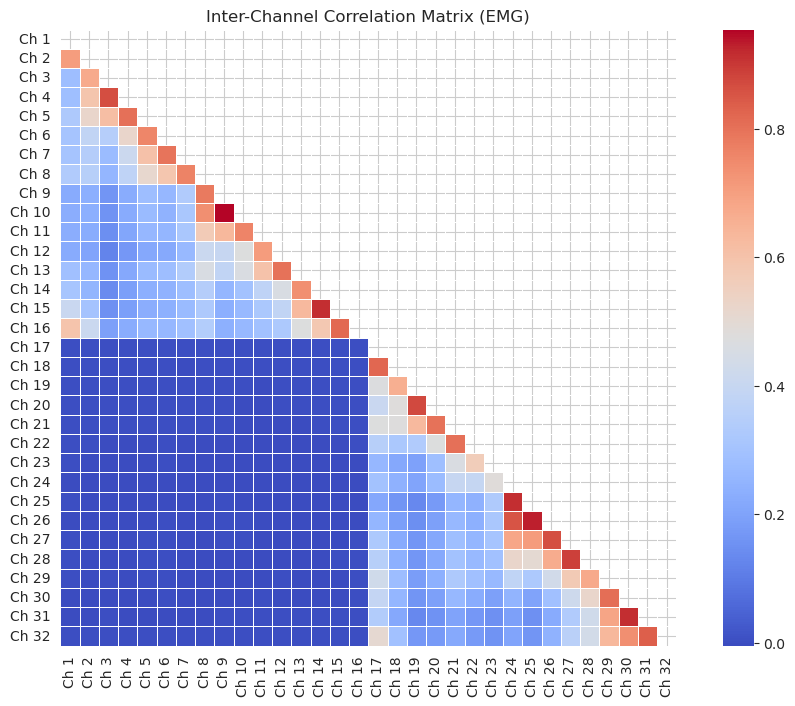}}
      \ffigbox[\FBwidth]{\caption{Log spectrogram of the signal}\label{fig:log}}{%
        \includegraphics[width=0.8\linewidth]{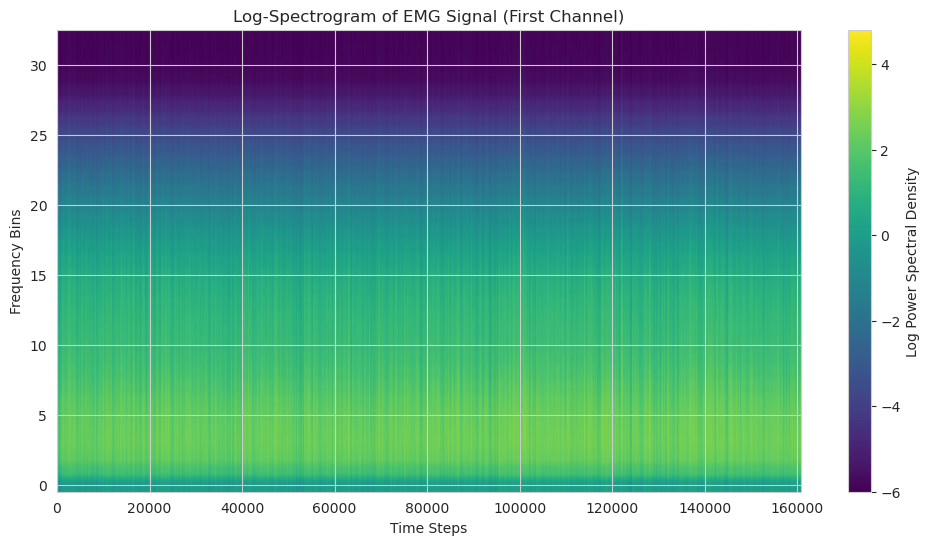}
      }
    \end{floatrow}
  \end{figure}

The dataset contains 32 sEMG input channels (16 per hand), distributed around the wrists. It is intuitive that channels positioned close together will exhibit correlated signals, while those farther apart (e.g., two or more channels away) should be largely independent. However, due to variability in wristband placement between users, slight misalignments can introduce unexpected correlations between more distant channels. To quantify this effect, we plotted the inter-channel correlation matrix, as shown in Fig.~\ref{fig:icc}. The results show that adjacent channels often display strong correlation, with influence extending up to two neighboring channels. This finding suggests the importance of incorporating rotational invariance into the model to ensure robustness across different users.

Moreover, the original emg2qwerty paper uses log-spectrogram representations as input to a Time-Delay Neural Network (TDS) encoder. We visualized several sample log-spectrograms (Fig.~\ref{fig:log}) and reviewed whether the chosen parameters adequately capture the relevant sEMG signal characteristics.

Based on our observations of the data, we applied several data augmentation techniques to enhance model generalization and training robustness. 

\begin{itemize}
\item \textbf{TemporalAlignmentJitter:} Introduces small temporal shifts to simulate misalignments caused by hardware latency, improving the model’s robustness to temporal variation.
\item \textbf{RandomBandRotation:} Randomly rotates EMG channel order to simulate variations in wristband placement across users, encouraging rotational invariance.
\item \textbf{SpecAugment:} Applies frequency masking, time masking, and time warping to the log-spectrograms, promoting generalization by preventing overfitting to specific temporal or spectral patterns.
\end{itemize}

\section{Methods}

In this section, we describe the baseline model provided by the emg2qwerty benchmark and outline our proposed extensions. Our goal is to build on the established TDS-based architecture while introducing attention-based encoders (Transformers and Conformers) to better capture the temporal and spatial structure of sEMG signals. We first summarize the baseline pipeline, then detail the modifications leading to our two proposed models.

\subsection{Baseline Model}

We adopt the official emg2qwerty baseline introduced by \citet{emg2qwerty}, the only existing architecture designed specifically for this dataset. The baseline follows an Automatic Speech Recognition (ASR)–inspired design: continuous multi-channel waveform inputs are mapped to character sequences through a log-spectrogram front-end and a Time-Depth Separable Convolution (TDS) encoder.

The dataset provides 32 channels of sEMG recorded at 2~kHz. These signals are converted into log-spectrogram features and normalized using 2D batch normalization across channels. To improve robustness across sessions and electrode placements, the baseline applies several augmentations: SpecAugment~\cite{park_19}, rotation augmentation (shifting channels by $-1,0,+1$), and temporal alignment jitter (0–120~samples) between the left and right hands. A small “Rotation-Invariance” MLP is applied per band to compensate for these shifts by averaging predictions across multiple rotated variants.

The core encoder is a stack of TDS blocks~\cite{hannun_19}, chosen for their ability to maintain wide receptive fields with relatively few parameters. The receptive field spans roughly 1~second of signal, capturing co-articulation effects that are prevalent in sEMG typing. The output sequence is trained using CTC loss~\cite{graves_06}, which outperforms cross-entropy in this setting. For decoding, the original emg2qwerty system integrates a 6-gram Kneser–Ney language model~\cite{heafield_13} into a beam-search decoder with custom backspace handling.

Finally, the baseline is acausal: it uses both past and future frames (approximately 900~ms history and 100~ms lookahead). While this improves accuracy, it is unsuitable for real-time use. Therefore, all of our proposed models adopt a strictly causal formulation using only past context.

\subsection{Proposed Models}

Our proposed models preserve the log-spectrogram front-end, normalization, and rotation-invariant MLP from the baseline, but replace or augment the TDS encoder with attention-based mechanisms. These encoders are responsible for extracting keystroke-relevant information while handling the complex temporal and cross-channel relationships inherent to sEMG.

The overall architecture is shown in Fig.~\ref{fig:architectures}. We evaluate two variants: a TDS–Transformer hybrid and a full Conformer encoder.

Transformers~\cite{transformer} have demonstrated strong performance across sequential tasks due to their ability to model long-range dependencies and global context. This is appealing for sEMG, where keystroke patterns depend on both local muscle activations and longer-term co-articulatory structure. The Conformer~\cite{conformer} augments Transformer layers with convolutional modules, making them especially suitable for physiological signals that exhibit strong local temporal structure alongside global dependencies.

\subsubsection{Approach 1) TDS - Transformer Encoder}
We added a Transformer Encoder layer in the baseline model. [Fig.~\ref{fig:architectures}] \citet{transformer} proposed the Transformer architecture, which relies entirely on self-attention mechanisms to model relationships in sequential data without using recurrence. This architecture has since become a standard in many ASR tasks due to its ability to capture long-range dependencies and contextual patterns effectively. 

To integrate the Transformer into our pipeline, we placed it after the convolutional layers that process the EMG signals. A linear projection layer was introduced to reduce the output feature dimension from the TDS Encoder by a factor of four (from 768 to 192) before feeding it into the Transformer. Since our ablation budget and resources were limited, we conducted a focused set of experiments that yielded several actionable insights. We experimented with various reduction sizes and found that this 4× compression offered the best performance, likely serving as a bottleneck that encourages the model to retain only the most salient information from the EMG signal. Additionally, while increasing the number of Transformer layers or attention heads did not result in further gains, we consistently observed that keeping the feedforward network (FFN) dimension above 2000 was beneficial, suggesting that a more expressive FFN helps capture the nonlinear structure of EMG data.

The addition of the Transformer Encoder consistently reduced CER in the generic model. Its self-attention mechanism enabled the model to focus on the most informative temporal regions, effectively capturing long-range dependencies within the EMG signal. This helped disambiguate similar patterns by leveraging contextual cues. 

\subsubsection{Approach 2) Conformer Encoder}

\begin{figure}[!ht]
    \centering
    \includegraphics[width=0.8\linewidth]{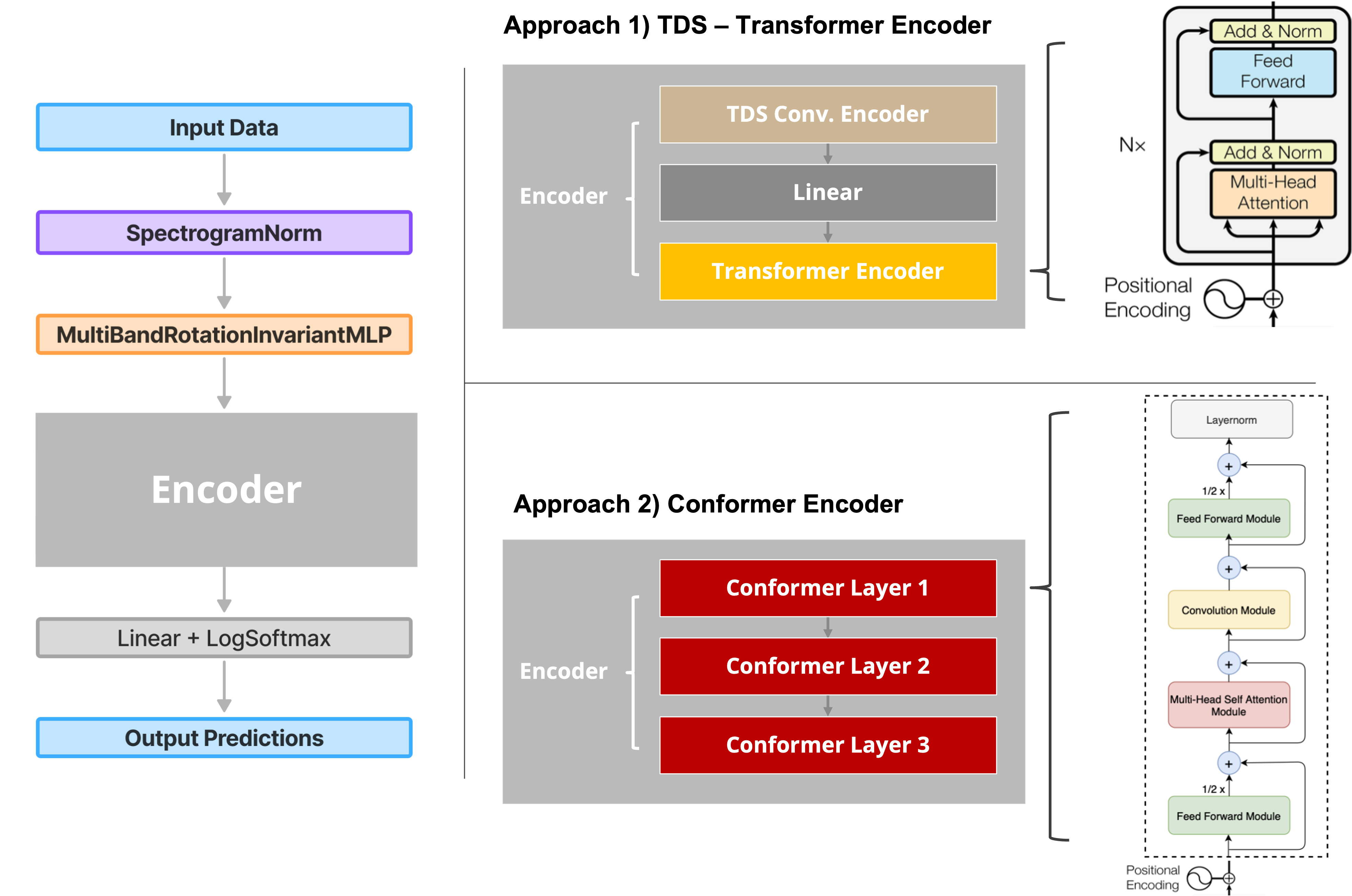}
    \caption{Proposed models diagram}
    \label{fig:architectures}
\end{figure}

As illustrated in Fig.~\ref{fig:architectures}, we adopt a 3-layer Conformer encoder with 6 attention heads in our model. In the original Conformer work \citet{conformer}, the authors propose a 16-layer, 4-head architecture tailored for Automatic Speech Recognition (ASR) tasks, where input sequences are typically long and consist of single-channel audio. In contrast, our sEMG dataset consists of relatively short sequences (4 seconds) with a much higher input dimensionality: 32 channels in total, corresponding to 16 sensors on each hand. Given this structural difference, directly applying the original configuration is neither efficient nor optimal.

To address this, we significantly reduce the number of layers in the encoder, as deeper networks are generally beneficial for capturing long-range temporal dependencies, which are less critical in our shorter input sequences. At the same time, we increase the number of attention heads to better leverage the multi-channel nature of the data. More attention heads allow the model to attend to interactions between different sEMG channels, which is essential for capturing the spatial relationships and cross-channel dynamics introduced by variability in sensor placement and muscle activation patterns. This adaptation ensures that the Conformer architecture is better aligned with the characteristics and demands of our physiological input modality.

\begin{table}[!ht]
\centering
\caption{Model architectures overview}
\label{tab:conformer-minimal}
\begin{tabular}{ll}
\toprule
\textbf{Approach} & \textbf{Details} \\
\midrule
TDS Encoder (Baseline) & $d_\text{in} = 528$, 4 blocks, 24 channels, kernel size = 32, MLP dim = 384 \\
TDS-Transformer Encoder & $d_\text{in} = 192$, 2 layers, 4 heads, kernel size = 32, FFN dim = 2048\\
Conformer Encoder & $d_\text{in} = 528$, 3 layers, 6 heads, kernel size = 31, FFN dim = 256 \\
\bottomrule
\end{tabular}
\end{table}

\subsection{Decoding Mechanism and LM}

\begin{figure}[H]
    \centering
    \includegraphics[width=1\linewidth]{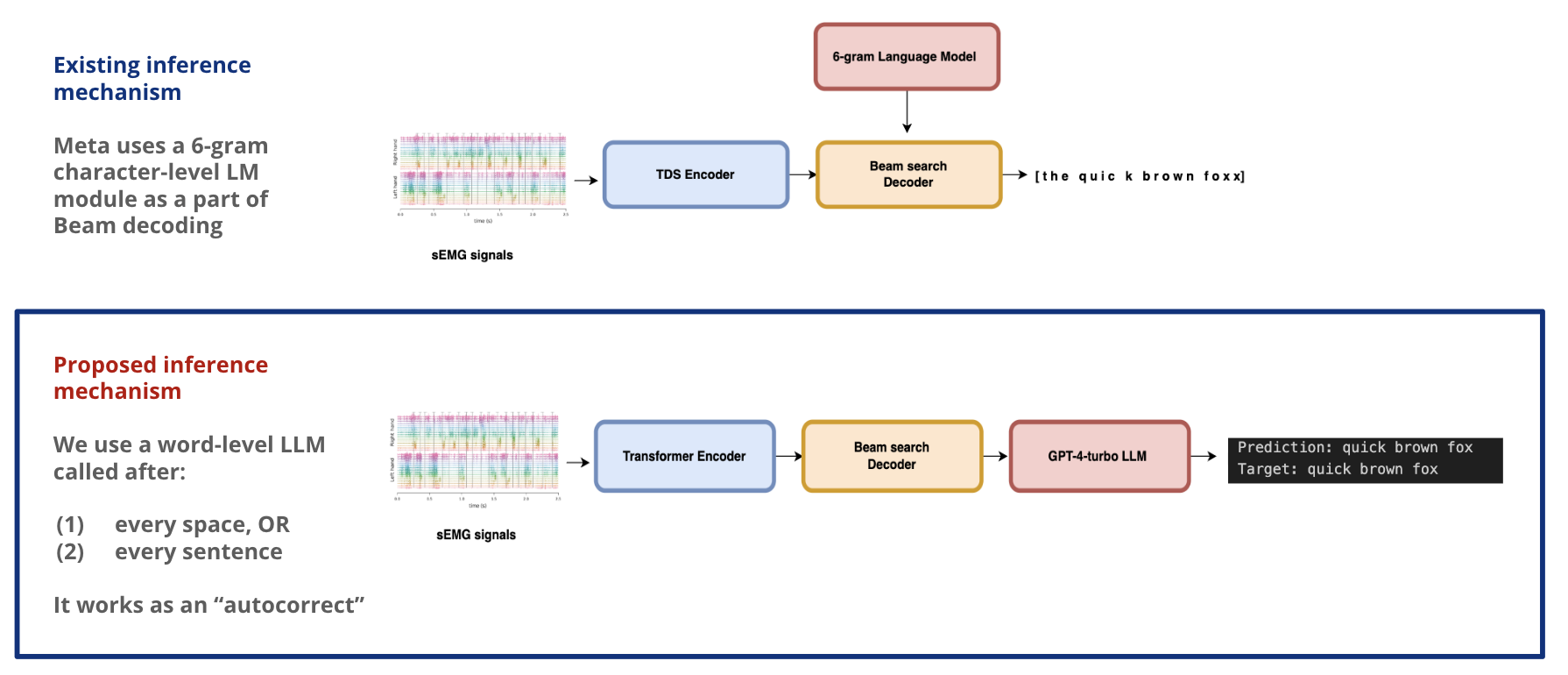}
    \caption{Proposed inference mechanism}
    \label{fig:lm}
\end{figure}

As shown in Fig.~\ref{fig:lm}, our decoding pipeline combines several strategies, augmented with language model (LM) capabilities, to refine the encoder’s raw predictions. This is particularly important in a virtual keyboard setting, where users type without physical keys or tactile feedback. Even experienced touch typists make more errors in such conditions, independent of model accuracy. The LM therefore serves as an intelligent auto-correction layer, improving overall reliability.

We evaluated multiple decoding approaches of increasing complexity. Greedy decoding serves as the baseline, selecting the most probable character at each timestep without considering alternative sequences. Beam search with an n-gram LM incorporates a character-level 6-gram language model that provides statistical priors over character sequences and maintains multiple hypotheses to improve global optimality.

For word-level LLM integration, we tested two variants:  
(1) space-level correction, in which the decoder invokes an LLM whenever a space is produced, enabling word-by-word refinement; and  
(2) sentence-level correction, where the decoder completes an entire sentence before invoking the LLM for a more global rewrite, reducing the number of API calls.

We experimented primarily with Flan-T5-small due to its favorable efficiency–performance trade-off for constrained deployments. We also tested larger models such as GPT-2 and GPT-4-turbo to assess the trade-offs between computational cost and correction quality. While these larger models offer improved accuracy and reasoning capabilities, they incur significantly higher latency and compute requirements.

\subsection{Loss Function and Evaluation Metrics}

We adopt the \textit{Connectionist Temporal Classification} (CTC) loss for training, as it is well-suited for sequence prediction tasks where the input and output sequences are aligned in order but not one-to-one in time. This is particularly appropriate for our task of mapping continuous sEMG signals to discrete keystroke sequences. CTC is especially compatible with transformer-based architectures and offers several advantages: (1) accommodates the variable-length nature of both the input sEMG signals and output keystroke sequences; (2) enables the model to learn alignments automatically, without requiring frame-level labels or precise timing information for each character, and (3) integrates smoothly with the transformer encoder output, allowing end-to-end training without additional alignment steps.

Formally, given an input sequence $\mathbf{x} = (x_1, x_2, \dots, x_T)$ of length $T$, and a target sequence $\mathbf{y} = (y_1, y_2, \dots, y_U)$ of length $U$, CTC defines the loss as the negative log-likelihood of the target sequence given all valid alignments $\mathcal{B}^{-1}(\mathbf{y})$:
\[
\mathcal{L}_{\text{CTC}} = -\log P(\mathbf{y} \mid \mathbf{x}) = -\log \sum_{\pi \in \mathcal{B}^{-1}(\mathbf{y})} P(\pi \mid \mathbf{x}),
\]
where $\pi$ is a path over the extended alphabet (including a blank symbol), and $\mathcal{B}$ is the collapsing function that removes repeated symbols and blanks from $\pi$ to produce the final output sequence $\mathbf{y}$.

For evaluation, we use the \textit{Character Error Rate} (CER) as the primary metric. CER is defined as the normalized Levenshtein distance (edit distance) between the predicted sequence $\hat{\mathbf{y}}$ and the ground truth $\mathbf{y}$:
\[
\text{CER} = \frac{S + D + I}{N},
\]
where:
\begin{itemize}
    \item $S$ = number of substitutions,
    \item $D$ = number of deletions,
    \item $I$ = number of insertions,
    \item $N$ = total number of characters in the ground truth sequence.
\end{itemize}

In addition to reporting the overall CER, we analyze its individual components—substitutions, insertions, and deletions—to gain deeper insight into the model’s behavior and identify common error modes.

\section{Results and Discussion}

\begin{table}[h]
\centering
\caption{
Combined CER results by model and decoding strategy. Greedy decoding is performed without a language model (LM); Beam decoding integrates external LMs (6-gram Char LM, Flan-T5, or GPT-4 Turbo). CER = Character Error Rate (\%). 
}
\label{tab:combined_model_decoding_results}
\resizebox{\textwidth}{!}{%
\begin{tabular}{|l|l|l|c|c|}
\toprule 
\textbf{Model} & \textbf{Decoder} & \textbf{LM} & \textbf{Generic CER (Online / Offline)} & \textbf{Personalized CER (\% Online / Offline)} \\
\midrule
TDS (baseline, acausal)          & Greedy        & None                           & -- / 55.38 & -- / 10.86 \\
TDS (baseline, causal)          & Greedy        & None                           & 24.98 / 58.32 & -- / -- \\
\midrule
TDS (baseline, causal)          & Beam          & 6-gram Char LM (baseline)      & -- / -- & -- / 45.55 \\
TDS (baseline, causal)          & Beam          & Flan-T5 (space-level)          & -- / -- & -- / 28.55 \\
TDS (baseline, causal)          & Beam          & Flan-T5 (sentence-level)       & -- / -- & -- / 42.65 \\
TDS (baseline, causal)          & Beam          & GPT-4 Turbo (space-level)      & -- / -- & -- / 28.11 \\
TDS (baseline, causal)          & Beam          & GPT-4 Turbo (sentence-level)   & -- / -- & -- / 27.65 \\
TDS (baseline, causal)          & Greedy        & Flan-T5 (space-level)          & -- / -- & -- / 28.55 \\
TDS (baseline, causal)          & Greedy        & Flan-T5 (sentence-level)       & -- / -- & -- / 42.65 \\
TDS (baseline, causal)          & Greedy        & GPT-4 Turbo (space-level)      & -- / -- & -- / 24.98 \\
TDS (baseline, causal)          & Greedy        & GPT-4 Turbo (sentence-level)   & -- / -- & -- / 22.71 \\
\midrule
\textbf{TDS + Transformer (causal)}  & Greedy        & None                           & \textbf{21.60 / 44.43} & -- / -- \\
\textbf{Conformer (causal)}             & Greedy        & None                           & \textbf{20.34} / \textbf{43.21}  & -- / \textbf{10.10} \\
\bottomrule
 \label{tab:results}
\end{tabular}
}
\end{table}

Table~\ref{tab:combined_model_decoding_results} summarizes all results across model architectures, decoding strategies, and language-model integrations. We first clarify the evaluation axes used throughout this section.

\textbf{Generic vs. personalized models.}  
Generic models are trained on the full user population and tested on unseen users, reflecting the real-world requirement of cross-user generalization. Personalized models are trained and evaluated on a single user and naturally achieve much lower error rates due to reduced inter-user variability in sEMG signals.

\textbf{Online vs. offline inference.}  
Online inference uses a sliding 4-second window to generate short, incremental outputs. Offline inference processes the entire sequence at once. 

\textbf{Causal vs. acausal modeling.}  
Acausal models can use both past and future context, while causal models operate strictly on past inputs and are suitable for real-time deployment. Despite the loss of future frames, the causal TDS baseline performs only slightly worse than the acausal version, likely because sEMG signals naturally precede physical motion by tens of milliseconds, providing inherent predictive information.

\medskip
\noindent\textbf{Baseline performance.}  
The original acausal TDS model achieves a strong personalized CER of 10.86\% but performs poorly in the generic offline setting (55.38\%). The causal TDS baseline reaches 24.98\% CER online and 58.32\% offline, confirming that user variability remains the dominant challenge for generic decoding.

\medskip
\noindent\textbf{Adding attention-based encoders: Transformer and Conformer.}  
Augmenting the TDS encoder with a Transformer layer yields clear improvements, reducing generic CER to 21.60\% online and 44.43\% offline. Replacing the encoder entirely with a 3-layer Conformer achieves the best generic results overall, lowering CER to 20.34\% (online) and 43.21\% (offline). Notably, the Conformer also achieves a strong personalized offline CER of 10.10\%, approaching the personalized acausal baseline while remaining fully causal.

\medskip
\noindent\textbf{Decoding strategies and language-model integration.}  
We also report personalized results with language-model–augmented decoding. Here, language models yield substantial gains. The best performance is achieved by GPT-4 Turbo with sentence-level correction on top of greedy decoding, reaching 22.71\% personalized offline CER. Flan-T5-small performs moderately well with space-level correction (28.55\%) but degrades under beam search, indicating limited ability to navigate larger hypothesis spaces.

\begin{figure}[h]
    \centering
    \includegraphics[width=0.3\linewidth]{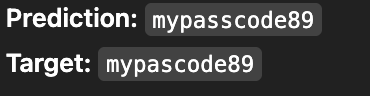}
    \caption{Sample prediction from gpt-4-turbo}
    \label{fig:pwdlm}
\end{figure}

Across all LLMs, beam search underperforms greedy decoding, suggesting that the added search complexity introduces misaligned hypotheses that even strong LLMs cannot reliably correct. This highlights a key trade-off in LLM-assisted typing: larger models improve fluency and correctness for natural text, but may inadvertently modify exact strings (e.g., ``usr123!'' → ``user123''), which is undesirable in contexts requiring literal reproduction, e.g., passwords or usernames, as illustrated in Fig.~\ref{fig:pwdlm}. 

\medskip
Overall, these results demonstrate a consistent trend:  
(1) attention-based encoders outperform convolutional baselines,  
(2) causal and online modeling is feasible with minimal degradation, and  
(3) lightweight LLM-based correction provides the strongest personalized accuracy gains.

\section{Conclusions}
We present a transformer-based sEMG decoding framework that substantially improves non-invasive muscle-driven typing. On the emg2qwerty dataset, attention-based encoders consistently outperform the TDS baseline, reducing online generic CER from 24.98\% → 20.34\% and offline personalized CER from 10.86\% → 10.10\%, We also convert the original acausal pipeline into a fully causal model with minimal accuracy loss, demonstrating its suitability for real-time use.

Latency remains the main obstacle, but the anticipatory nature of sEMG signals and predictive inference suggest that sub–100~ms performance is achievable. Combined with lightweight LLM-based decoding, these results highlight the feasibility of accurate, real-time, hands-free text input.

Overall, our findings position sEMG as a promising, natural input modality for spatial, wearable, and screenless computing, and point toward muscle-based interaction as a key building block of future HCI systems.

\section{Future Work}

The most significant challenge is inference latency. Under our current online inference setup, users must wait 4 seconds before the first keystroke appears on the screen (this delay is clearly unsuitable for real-world applications). Ideally, the end-to-end latency should be under 50 ms, which matches the response time of a typical Bluetooth keyboard.

One potential solution is to drastically reduce the sliding window size while maintaining a longer context window through padding. For instance, using a 50 ms sliding window with a 4-second historical padding window would allow the model to retain the same input size while enabling real-time updates. In our current setup, each inference takes roughly 50 ms, which would result in an effective latency of approximately 100 ms. Though still higher than hardware latency, this is a meaningful step forward.

Moreover, we can leverage domain knowledge in EMG, where muscle activation typically precedes physical movement by tens of milliseconds. This opens the door to predictive modeling, where the system anticipates keystrokes before they occur, potentially achieving less than 50 ms latency. That said, even our personalized models currently yield around 10\% CER, which suggests the technology is not yet ready for deployment in real-world scenarios. Nonetheless, reducing latency and increasing accuracy are critical next steps toward practical, muscle-driven typing systems.

\section*{Acknowledgment}
We would like to thank Professor Bhiksha Raj of Carnegie Mellon University for his guidance and support throughout this project.

This work was conducted as part of the Carnegie Mellon University 11-785 Deep Learning course (Spring 2025):  
\href{https://deeplearning.cs.cmu.edu/S25/index.html}{https://deeplearning.cs.cmu.edu/S25/index.html}


\section*{Code}
The code for our experiments is available at:  
\url{https://github.com/KunwooLeeKay/emg2qwerty}


\bibliography{refs}


\end{document}